\documentclass[sigconf, 10pt]{acmart}

\usepackage{booktabs} % For formal tables

\setcopyright{acmcopyright}
\usepackage{graphicx}
\usepackage{balance}
\usepackage{float}
\usepackage{amssymb}
\usepackage{multirow}
\usepackage{array}
\usepackage{subfigure}
\usepackage{color}
\usepackage[ruled,linesnumbered,boxed]{algorithm2e}
\usepackage{color}

\newtheorem{definition}{Definition}

\newtheorem{example}{Example}

\acmDOI{10.1145/3242153.3242156}

\acmISBN{978-1-4503-6607-6/18/08}

\acmConference[BIRTE'18]{International Workshop on Real-Time Business Intelligence and Analytics}{August 27, 2018}{Rio de Janeiro, Brazil}
\acmYear{2018}
\copyrightyear{2018}

\acmPrice{15.00}

\acmBooktitle{International Workshop on Real-Time Business Intelligence and Analytics (BIRTE '18), August 27, 2018, Rio de Janeiro, Brazil}
\begin{document}
\title[Improving Machine-based Entity Resolution \\ with Limited Human Effort: A Risk Perspective]{Improving Machine-based Entity Resolution with Limited Human Effort: A Risk Perspective}
\author{Zhaoqiang Chen, Qun Chen, Boyi Hou, Murtadha Ahmed, Zhanhuai Li}
\affiliation{%
	\vspace{-8pt}
  $^1$School of Computer Science, Northwestern Polytechnical University, China \\
  $^2$Key Laboratory of Big Data Storage and Management, NPU, MIIT, China%\\
  %1 Dongxiang Road, Xi'an Shaanxi, 710129, China
}
\email{{chenzhaoqiang@mail., chenbenben@, ntoskrnl@mail., a.murtadha@mail., lizhh@}nwpu.edu.cn}

\renewcommand{\shortauthors}{Z. Chen, Q. Chen et al.}

\begin{abstract}
 Pure machine-based solutions usually struggle in the challenging classification tasks such as entity resolution (ER). To alleviate this problem, a recent trend is to involve the human in the resolution process, most notably the crowdsourcing approach. However, it remains very challenging to effectively improve machine-based entity resolution with limited human effort. In this paper, we investigate the problem of human and machine cooperation for ER from a risk perspective. We propose to select the machine-labeled instances at high risk of being mislabeled for manual verification. For this task, we present a risk model that takes into consideration the human-labeled instances as well as the output of machine resolution. Finally, we evaluate the performance of the proposed risk model on real data. Our experiments demonstrate that it can pick up the mislabeled instances with considerably higher accuracy than the existing alternatives. Provided with the same amount of human cost budget, it can also achieve better resolution quality than the state-of-the-art approach based on active learning. \vspace{-6pt}
\end{abstract}





\maketitle

\vspace{-10pt} \section{introduction}
	Entity resolution aims at finding the records that refer to the same real-world entity. Usually considered as a classification task, ER is challenging in that the records may contain incomplete and dirty values. ER can be performed based on rules, probabilistic theory or machine learning \cite{Singh2017, christen2012data}. However, the traditional machine-based solutions may not be able to produce satisfactory results in many practical scenarios. Therefore, there is an increasing need to involve the human in the resolution process for improved quality \cite{wang2012crowder}. For instance, the active learning approach \cite{sarawagi2002interactive} proposed to select the instances for manual verification based on the benefit they can bring to a machine classifier. The approach of crowdsourcing \cite{Jain2017, wang2012crowder} instead investigated how to make the human work efficiently and effectively on a given workload. Depending on pre-specified assumptions (e.g. partial order relationship \cite{chai2016cost}), it usually makes the human label some instances in a workload for the purpose that the remaining instances can be automatically labeled by the machine with high accuracy.
	
	It can be observed that the existing hybrid approaches select the instances for manual verification to maximize the benefit they can bring to a given workload as a whole. However, the marginal benefit of additional manual work usually decreases (sometimes dramatically) with the cost. For instance, in active learning, it has been well recognized \cite{schohn2000less} that increasing the number of training data points may quickly become ineffectual in improving classification performance after initial iterations. In the application scenarios where fast response is required, it is also desirable that a limited amount of human effort can be exclusively spent on the instances at high risk of being mislabeled by the machine.
	
	In this paper, we investigate the problem of human and machine cooperation for improved quality from a risk perspective. Given a limited human cost budget, we propose to select the machine-labeled instances at high risk of being mislabeled for manual verification. The proposed risk-based solution is supposed to be used in the scenario where increasing training points for a learning model has become ineffectual or not cost-effective in improving classification performance. It can therefore serve as a valuable complement to the existing learning-based solutions. On the other hand, even though some of the proposed techniques for active learning (e.g. training instance selection based on uncertainty \cite{mozafari2014scaling}) can be naturally applied for this task, our work is the first to introduce the concept of risk and propose a formal risk model for the task. The major contributions of this paper can be summarized as follows:

\begin{itemize}
\item We investigate the problem of human and machine cooperation for ER from a risk perspective and define the corresponding optimization problem (Section.~\ref{sec:problem});
\item We present a risk model for prioritizing the machine-labeled instances for manual verification (Section.~\ref{sec:riskmodel});
\item We evaluate the performance of the proposed approach on real data by a comparative study. The experimental results validate its efficacy (Section.~\ref{sec:experiment}).
\end{itemize}

\vspace{-3pt} \section{Problem Definition} \label{sec:problem}
   Given an ER workload consisting of record pairs, a machine classifier labels each pair as {\em match} or {\em unmatch}. Due to the inherent challenge of entity resolution, a classifier may be prone to mislabeling some of the pairs. In this paper, we investigate the problem of how to improve the results of machine resolution by manually correcting machine errors. Since human work is expensive, we impose a budget on the amount of spent human effort. For the sake of presentation simplicity, we quantify the budget by the number of manually-inspected pairs. Given a budget $k$, an ideal solution would identify $k$ mislabeled pairs. In this case, each manual inspection effectively corrects a machine error. However, in practice, it is more likely that a solution chooses both mislabeled and correctly labeled pairs. We formally define the optimization problem as follows:
\begin{definition}
\label{problemdefinition}
  {\bf [Optimization Problem of Improving Machine Resolution by Manual Inspection].} Given an ER workload, $D$, which consists of $n$ record pairs, \{$d_1$,$d_2$,$\ldots$,$d_n$\}, a machine classifier labels each pair in $D$ as {\em match} or {\em unmatch}. Given the budget $k$ on human work, the optimization problem is to identify a set of $k$ machine-labeled pairs in $D$, denoted by $D_H$, for manual inspection such that the number of pairs misclassified by the machine in $D_H$ is maximized.
\end{definition}		

\begin{figure}
\centering
\includegraphics[width=0.9\linewidth]{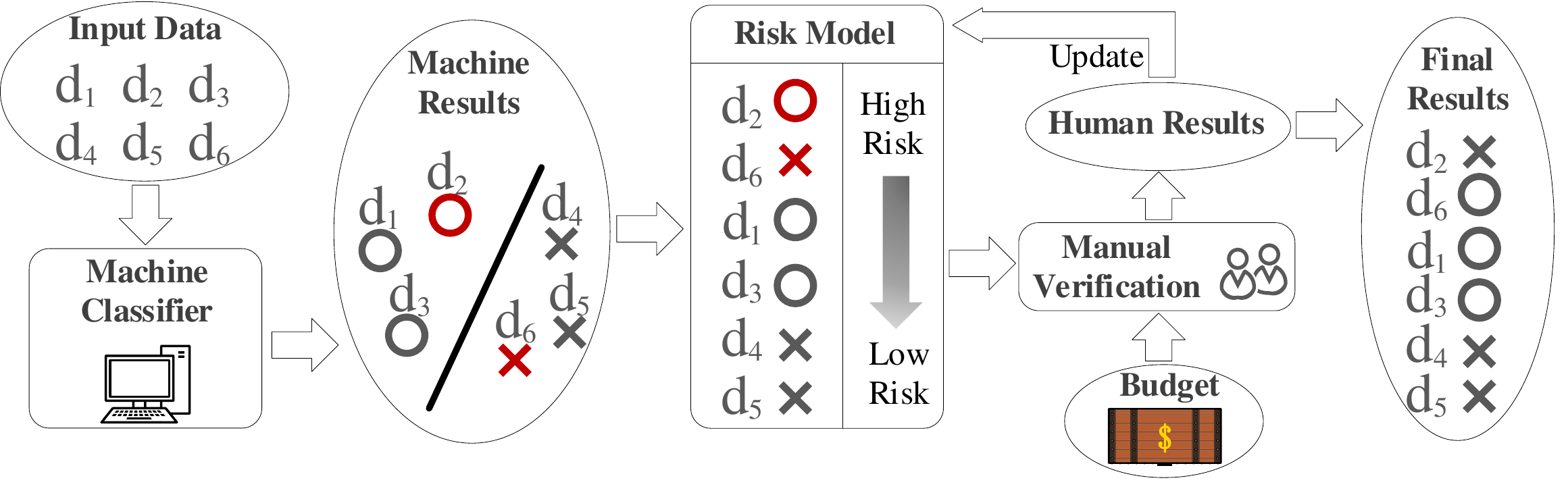}
\vspace{-0.3cm}
\caption{The Risk-based Solution.}
\label{fig:overview}
\vspace{-0.4cm}
\end{figure}

\hspace{-0.14in}{\bf Risk-based Solution.}  The optimization problem defined in Definition.~\ref{problemdefinition} is challenging due to the fact that the match probabilities of the machine-labeled pairs are difficult to estimate.  In this paper, we propose to solve the optimization problem from a risk perspective. In other words, the machine-labeled pairs at higher risk of being mislabeled should be chosen first for manual inspection. It can be observed that if risk measurement is accurate given all the available information, the strategy of selecting by risk-wise order can be considered optimal. The workflow of the risk-based solution is presented in Figure.~\ref{fig:overview}. It iteratively selects the most risky machine-labeled pairs for manual inspection until the budget limit is reached. After each iteration, the set of manually-labeled pairs is updated, and is used to re-evaluate the risk of the remaining machine-labeled pairs.

  It is worthy to point out that the risk-based solution can work properly with both supervised and unsupervised classifiers. Given a supervised classifier, risk analysis can be initially performed based on the human-labeled pairs as well as machine resolution. Given an unsupervised classifier, risk analysis can only start with machine resolution; after initial iterations, it can then be similarly performed based on the human-labeled pairs as well as machine resolution.

\vspace{-3pt} \section{Risk Analysis} \label{sec:riskmodel}
   In this section, we propose the technique of risk analysis for prioritizing pair selection. Given an instance pair $d_i$ in $D$, we represent its match probability by a random variable, $P_i$. As usual, we model $P_i$ by a normal distribution, $\mathcal{N}(\mu_i, \sigma_i^2)$, where $\mu_i$ and $\sigma_i^2$ denote its expectation and variance respectively. In the rest of this section, we first describe how to estimate the match probability distribution in Subsection~\ref{sec:bayesian-analysis}, and then present the metric for risk measurement in Subsection~\ref{sec:cvar}.

\vspace{-3pt} \subsection{Distribution Estimation} \label{sec:bayesian-analysis}
  It can be observed that there exist two information sources for the estimation of match probability distribution. Firstly, even though a machine classifier may fail to produce satisfactory resolution results, it can provide valuable hints about the status of the pairs. Therefore, the results of machine resolution can generally serve as a starting point for the estimation. The second source consists of the human-labeled results. Compared with machine labels, the labels provided by the human are usually more accurate, i.e. they can provide more information beyond the capability of machine resolution.
	
	We employ the classical Bayesian inference \cite{berger1985statistical} to estimate the distribution. The inference process takes the match probability estimated by the machine as the prior expectation, and uses the human-labeled pairs as samples to estimate the posterior expectation and variance. The proposed approach has the desirable property that it can seamlessly integrate the hints provided by both the human and the machine into a unified inference process.

\vspace{-3pt}  \subsubsection{Prior expectation estimation by machine}
   A machine classifier labels instance pairs as match or unmatch based on a classification metric. Generally, the match probability of a pair can be considered to be monotonous with its metric value. In this paper, we use the SVM (Support Vector Machine) classifier based on active learning as the illustrative example. It classifies pairs through a hyperplane. Instead of randomly selecting training data points, it iteratively chooses the instance pair which is closest to the hyperplane of the current SVM as the next training data point, and updates the SVM until a preset training budget is exhausted. Note that an SVM classifier usually provides a pair's distance from the hyperplane, rather than a match probability, as the evidence for its given label. We therefore use Platt's probabilistic outputs for SVM \cite{platt1999probabilistic} to translate the distance into a match probability.

\vspace{-3pt} \subsubsection{Sample observation generation by human}
	We generate the sample observations on the status of a target pair based on features. Features serve as the medium to convey valuable information from the human-labeled pairs to a target pair. Desirably, the features used for information conveyance should have the following three properties: 
\begin{enumerate}	
\item They can be easily extracted from the human-labeled pairs; 
\item They should be evidential, or indicative of the status of a pair; 
\item They should be to a large extent independent of the metric used by the machine classifier. 
\end{enumerate}
The final property ensures that the sample observations can provide additional valuable information not implied by machine labels. To this aim, we extract two types of features from pairs, \emph{Same($t_i$)} and \emph{Diff($t_i$)}, where $t_i$ represents a token, \emph{Same($t_i$)} indicates that both records in a pair contain $t_i$, and \emph{Diff($t_i$)} indicates that one and only one record in a pair contains $t_i$.  It can be observed that these two features are evidential and easily extractable. Moreover, they were not used in the existing classification metrics proposed for ER.
	
	Suppose that a target pair, $d_i$, contains $m$ features, which are denoted by \{$f_1$, $f_2$, $\ldots$, $f_m$\}. A human-labeled pair containing all the $m$ features can be naturally considered to be a valid observation on the status of $d_i$. Unfortunately, due to their limited number in practical scenarios, the human-labeled pairs with this property may not provide with sufficient observations. Therefore, we also consider the human-labeled pairs that contain only a portion of the $m$ features in $d_i$. Suppose that a human-labeled pair, $d_j^h$, contains the $k$ features in $d_i$, \{$f_1$, $f_2$, $\ldots$, $f_k$\}, but does not contain the remaining $(m-k)$ features. Inspired by the portfolio investment theory \cite{rockafellar2002conditional}, we treat features as stocks, and a feature's match probability as its investment reward. Then, the match probability of $d_i$ corresponds to the combined reward of an investment portfolio consisting of $m$ stocks, \{$f_1$, $f_2$, $\ldots$, $f_m$\}. 
	
	Based on the label of $d_j^h$, we generate the corresponding sample observation on the status of $d_i$ by
\begin{equation}
   O_j(d_i)=\frac{L(d_j^h)+\sum_{k<r\leq m}{w_rE(f_r)}}{1 + \sum_{k<r\leq m}{w_r}},
\label{eq:observation}
\end{equation}		
in which $w_r$ denotes the feature weight, $L(d_j^h)$ denotes the manual label of $d_j^h$, and $E(f_r)$ denotes the expectation of $f_r$'s match probability. In Eq.~\ref{eq:observation}, $L(d_j^h)$=1 if the label is {\em match} and $L(d_j^h)$=0 otherwise. We estimate $E(f_r)$ by
\begin{equation}
  E(f_r)=\frac{\sum_{1\leq s\leq n}{L(d_s^r)}}{n},
\end{equation}
in which $d_s^r$ denotes a human-labeled pair containing the feature $f_r$ and $n$ denotes its total number. An example of sample observation generation is shown in Example~\ref{exam:observation-generation}. More details can be found in our technical report \cite{chen2018riskerreport}. It is worthy to point out that in the generation of sample observations for $d_i$, we only consider the features contained in the human-labeled pairs.  If a feature of $d_i$ never appears in the human-labeled pairs, we lack reliable information to reason about its match probability. It is therefore ignored in the observation generation process.

\begin{example} \label{exam:observation-generation}
  Suppose that a target pair, $d_1$, contains 3 features, \{$f_1$,$f_2$,$f_3$\}, and a pair manually labeled as {\em unmatch} by the human, $d_2^h$, contains $f_1$ and $f_2$, but not $f_3$. For the sake of presentation simplicity, we also suppose that feature weights are equally set to be 1.
With the expectation of the match probability of $f_3$ being estimated at 0.3, the sample observation provided by $d_2^h$ for the status of $d_1$ is approximated by $O_2(d_1)=\frac{0+0.3}{2}=0.15$.
\end{example}

\vspace{-3pt} \subsubsection{Bayesian inference}
	 Given a random variable $V$ following a known prior distribution, $\pi(V)$, the technique of Bayesian inference ~\cite{berger1985statistical} estimates the posterior distribution of $V$ by combining the prior information provided by $\pi(V)$ and the sample observations. In our example, the prior distribution of the match probability of a target pair, $d_i$, is represented by the normal distribution of $\mathcal{N}(\mu_i, \sigma_i^2)$. Suppose that the prior expectation of $\mu_i$ provided by the machine classifier is $\mu_i^0$ and the human-labeled pairs provide with $n$ sample observations.

  As usual, we suppose that $\mu_i$ and $\sigma_i^2$ follow a combined conjugate prior distribution, or a {\em normal-inverse-gamma distribution}. The prior distributions of $\mu_i$ and $\sigma_i^2$ can thus be represented by
\begin{equation}
p(\mu_i|\sigma_i^2; \mu_i^0, n^0) \sim \mathcal{N}(\mu_i^0, \frac{\sigma_i^2}{n^0}),
\label{eq:mean-dist}
\end{equation}
and
\begin{equation}
p(\sigma_i^2; \alpha, \beta) \sim InvGamma(\alpha^0, \beta^0),
\label{eq:variance-dist}
\end{equation}
where $n^0$, $\alpha^0$ and $\beta^0$ are the hyperparameters, and $InvGamma()$ denotes an {\em inverse-gamma distribution}. Denoting the posteriors by $\mathcal{N}(\mu_i^1, \frac{\sigma_i^2}{n^1})$ and $InvGamma(\alpha^1, \beta^1)$, we have
\begin{equation}
\begin{split}
&\mu_i^1 = \frac{n^0\cdot\mu_i^0 + n\cdot \bar{p}_i}{n^0 + n}, \\
&n^1 = n^0 + n, \\
&\alpha^1 = \alpha^0 + \frac{n}{2}, \\
\beta^1 = \beta^0 + \frac{1}{2}\sum&_{j=1}^{n}(p_i^j - \bar{p}_i)^2 + \frac{1}{2}\cdot \frac{n^0 n}{n^0 + n}\cdot (\mu_i^0 - \bar{p}_i)^2,
\end{split}
\end{equation}
where $\bar{p}_i$ denotes the average value of observed samples. 

  In Eq.~\ref{eq:mean-dist} and ~\ref{eq:variance-dist}, the hyperparameters $n^0$, $\alpha^0$ and $\beta^0$ are used to convey the belief about the prior information. Specifically, given a confidence level of $\theta$ on the prior expectation $\mu_i^0$, we set $n^0=\theta n / (1 - \theta)$. It means that the inference process will preserve $\theta \mu_i^0$ for the estimation of $\mu_i$. Similarly, we set $\alpha^0=\frac{n}{2}\cdot\frac{\theta}{1-\theta} + 1$, and $\beta^0 = S_n^2 \cdot (\alpha^0 - 1)$, in which $S_n^2$ represents the variance of all the samples. It means that the inference process will preserve $\theta S_n^2$ for the estimation of $\sigma_i^2$.
	
	Based on the obtained posterior distributions of $\mu_i$ and $\sigma_i^2$, a point estimate $\hat{\mu}_i$ for the random variable $\mu_i$ (resp. $\hat{\sigma}_i^2$ for $\sigma_i^2$) can be inferred using a metric of Bayes risk. More details on the Bayesian inference can be found in our technical report~\cite{chen2018riskerreport}. 
		
\vspace{-3pt} \subsection{Risk Model} \label{sec:cvar}
    Inspired by the portfolio investment theory~\cite{rockafellar2002conditional}, we employ the metric of Conditional Value at Risk (CVaR) to measure the risk of pairs being mislabeled by the machine. Given a confidence level of $\theta$, CVaR is the expected loss incurred in the $1-\theta$ worst cases. Formally, given the loss function $z(X)\in L^p(\mathcal{F})$ of a portfolio $X$ and $\theta$, the metric of CVaR is defined as follows:
\begin{equation}
CVaR_\theta(X) = \frac{1}{1-\theta}\int_0^{1-\theta}VaR_{1-\gamma}(X)d\gamma,
\end{equation}
where $VaR_{1-\gamma}(X)$ represents the minimum loss incurred at or below $\gamma$ and can be formally represented by
\begin{equation}
VaR_{1-\gamma}(X)=inf\{z_*: P(z(X)\geq z_*)\leq \gamma\}.
\end{equation}

  Given a pair, $d_i$, we denote its match probability by $x$, and its probability density function and cumulative distribution function by $pdf_{d_i}(x)$ and $cdf_{d_i}(x)$ respectively. If $d_i$ is labeled by the machine as {\em unmatch}, its probability of being mislabeled by the machine is equal to $x$. Accordingly, its worst-case loss corresponds to the case that $x$ is maximal. Therefore, given the confidence level of $\theta$, the CVaR of $d_i$ is the expectation of $z=x$ in the $1-\theta$ cases where $x$ is from $cdf_{d_i}^{ - 1}(\theta)$ to $+ \infty$. Formally, the CVaR risk of a pair $d_i$ with the machine label of {\em unmatch} can be estimated by
\begin{equation}
CVa{R_\theta }(d_i) =
\frac{1}{1 - \theta }\int\limits_{{{cdf}_{d_i}}^{ - 1}(\theta )}^{ + \infty } {pdf_{d_i}} (x) \cdot xdx.
\end{equation}

Otherwise, if $d_i$ is labeled by the machine as {\em match}, its potential loss of being mislabeled by the machine is equal to 1-$x$. Therefore, the CVaR risk of a pair $d_i$ with the machine label of {\em match} can be similarly estimated by
\begin{equation}
CVa{R_\theta }(d_i) =
\frac{1}{1 - \theta}\int\limits_{ - \infty }^{{cdf_{d_i}}^{ - 1}(1 - \theta )} {pdf_{d_i}} (x) \cdot (1 - x)dx.
\end{equation}
\vspace{-3pt} \section{Empirical Evaluation} \label{sec:experiment}
    We have evaluated the performance of the proposed risk model, denoted by CVAR, on real data by a comparative study. We compare it with both a baseline alternative and a state-of-the-art technique proposed for active learning \cite{mozafari2014scaling}. The baseline method, denoted by BASE, selects the machine-labeled pairs solely based on the match expectation estimated by the machine. Specifically, given a pair $d_i$ and its match probability $\mu_i^0$ provided by a classifier, the risk of $d_i$ with the machine label of {\em unmatch} (resp. {\em match}) is simply estimated to be $\mu_i^0$ (resp. ($1-\mu_i^0$)). Since the two algorithms proposed in \cite{mozafari2014scaling}, {\em Uncertainty} and {\em MinExpError}, perform very similarly in our experiments, we only report the results of {\em Uncertainty}. We denote the algorithm of {\em Uncertainty} by UNCT. Intuitively,  UNCT iteratively selects the pairs that the classifier is most uncertain about for manual verification. 
    
    Additionally, we also compare the proposed risk-based solution (denoted by RISK) with the active learning solution (denoted by ACTL) on the achieved resolution quality provided with the same amount of human cost budget. Note that the ACTL solution would tune classifier parameters after additional manual verification, thus can potentially improve classification accuracy, while RISK would not.

    We used the real datasets DBLP-Scholar~\footnote{\url{https://dbs.uni-leipzig.de/file/DBLP-Scholar.zip}} and Abt-Buy~\footnote{\url{https://dbs.uni-leipzig.de/file/Abt-Buy.zip}} in the empirical study.  As usual, we use the standard blocking technique to filter the instance pairs unlikely to match. After blocking, the DBLP-Scholar workload contains totally $41416$ instance pairs, and the Abt-Buy workload contains totally $20314$ instance pairs. We employ SVM as the machine classifier. On DBLP-Scholar, we use the Jaccard similarity over the attributes {\em title} and {\em authors}, the edit distance over the attributes {\em title}, {\em authors} and {\em venue}, and the number equality over {\em publication year} as the input features for SVM. With only $1\%$ of input data as training data, the achieved precision and recall of the SVM classifier are $0.917$ and $0.875$ respectively. On Abt-Buy, we use the Jaccard similarity and edit distance over the attributes {\em product name} and {\em description} respectively as the input features for SVM. With only 2\% of input data as training data, the achieved precision and recall are $0.567$ and $0.338$ respectively. In the implementation of risk analysis, the confidence level $\theta$ is set to 0.8. Since a valid match probability should be between 0 and 1, we transform the inferred normal distribution to a {\em truncated normal distribution} in the range of 0 to 1 \cite{burkardt2014truncated}.

\begin{figure}
\centering
\subfigure[The DBLP-Scholar dataset.]
{\includegraphics[width=0.45\linewidth]{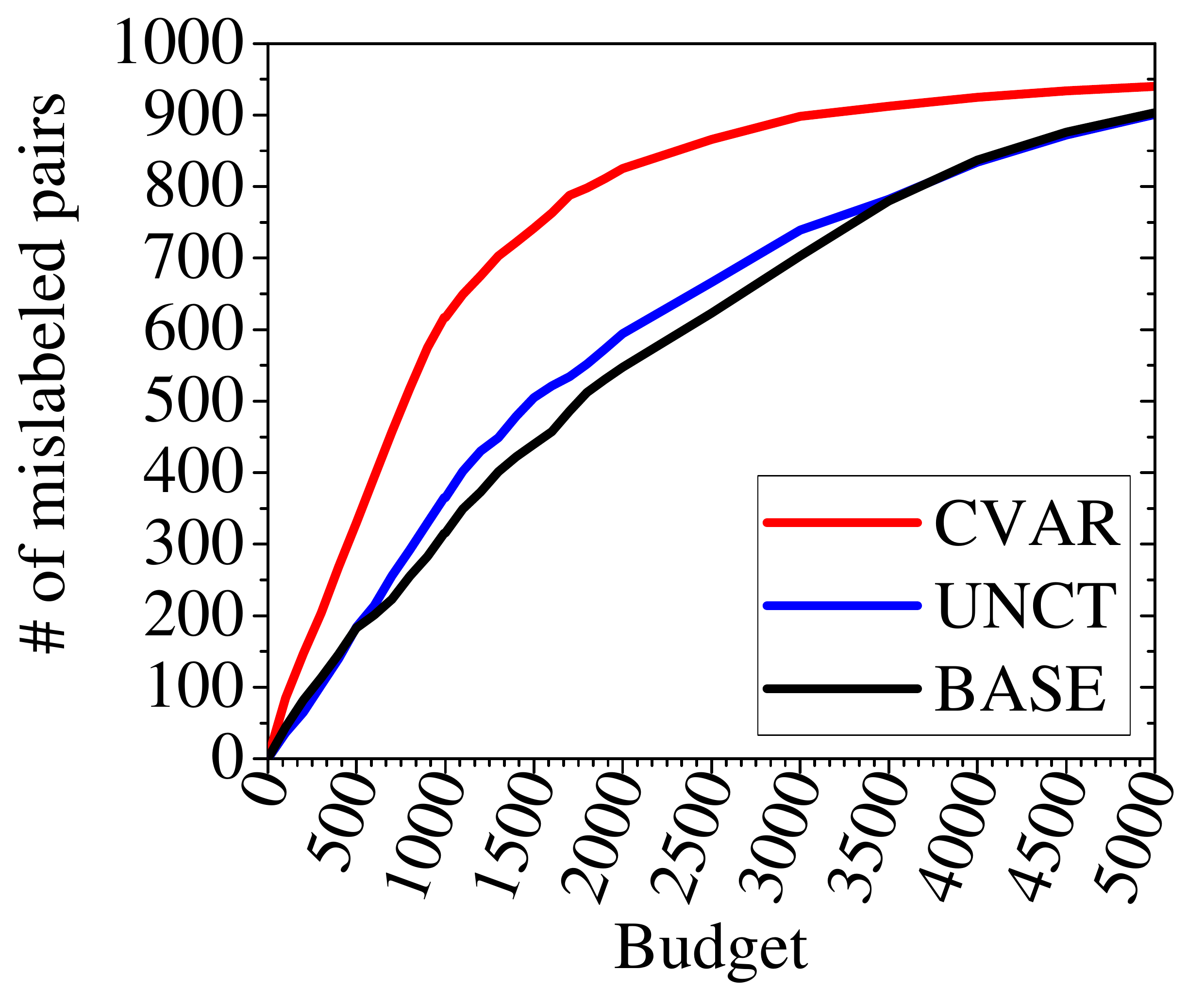}}
\subfigure[The Abt-Buy dataset.]
{\includegraphics[width=0.45\linewidth]{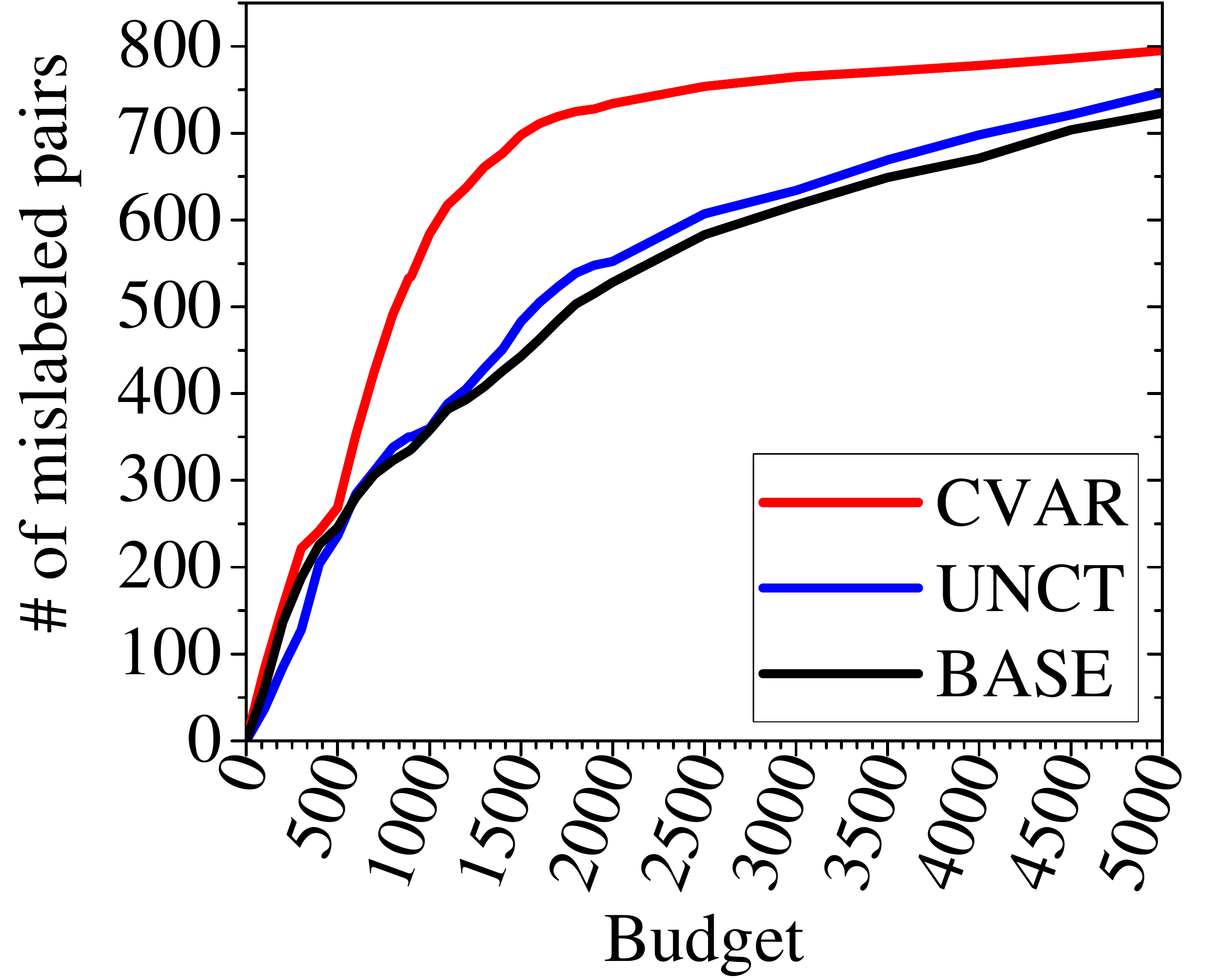}}
\vspace{-0.5cm}
\caption{Pick-up accuracy comparison.}
\label{fig:pairs-number}
\vspace{-0.5cm}
\end{figure}

\begin{figure}
\centering
\subfigure[The DBLP-Scholar dataset.]
{\includegraphics[width=0.45\linewidth]{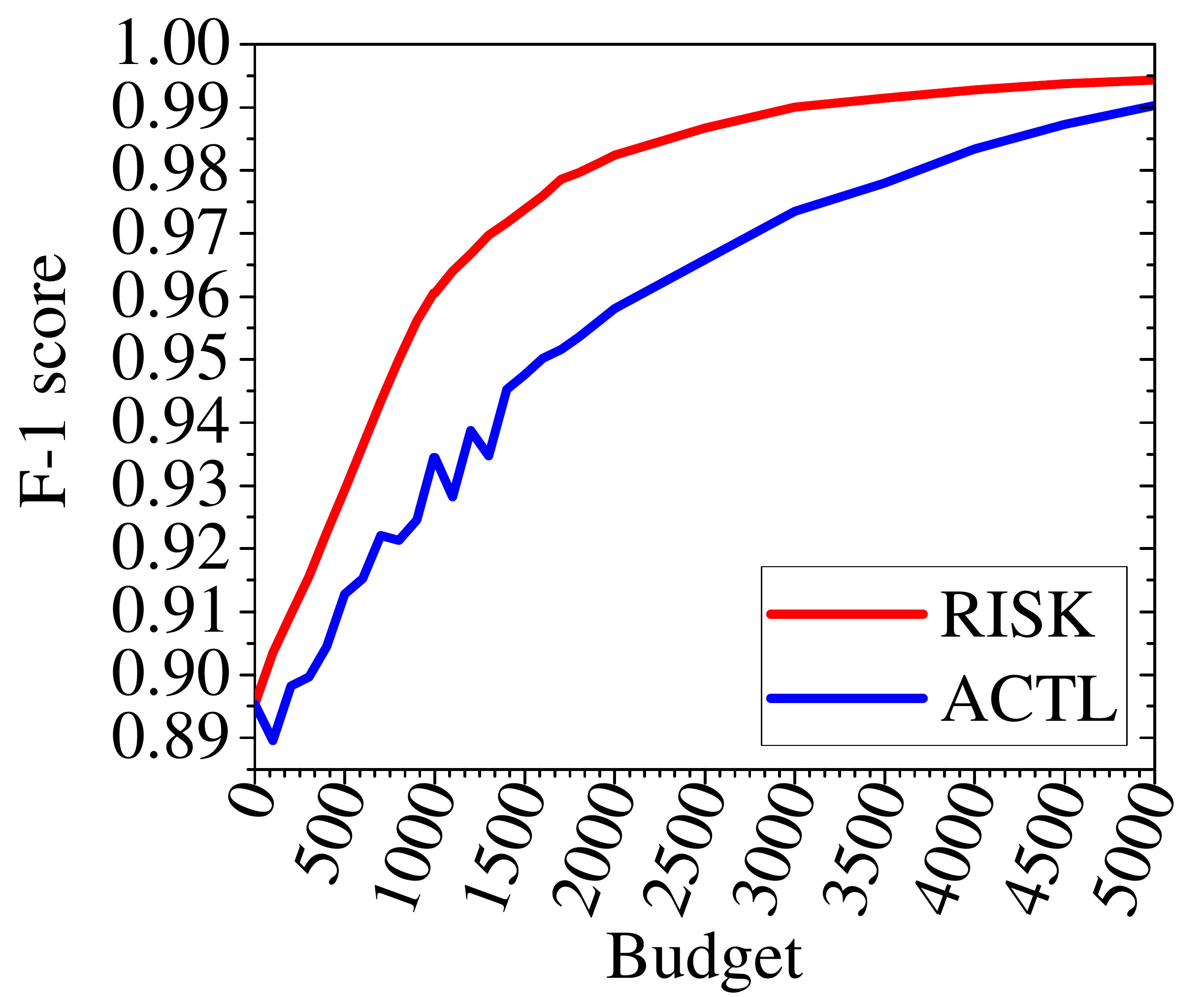}}
\subfigure[The Abt-Buy dataset.]
{\includegraphics[width=0.45\linewidth]{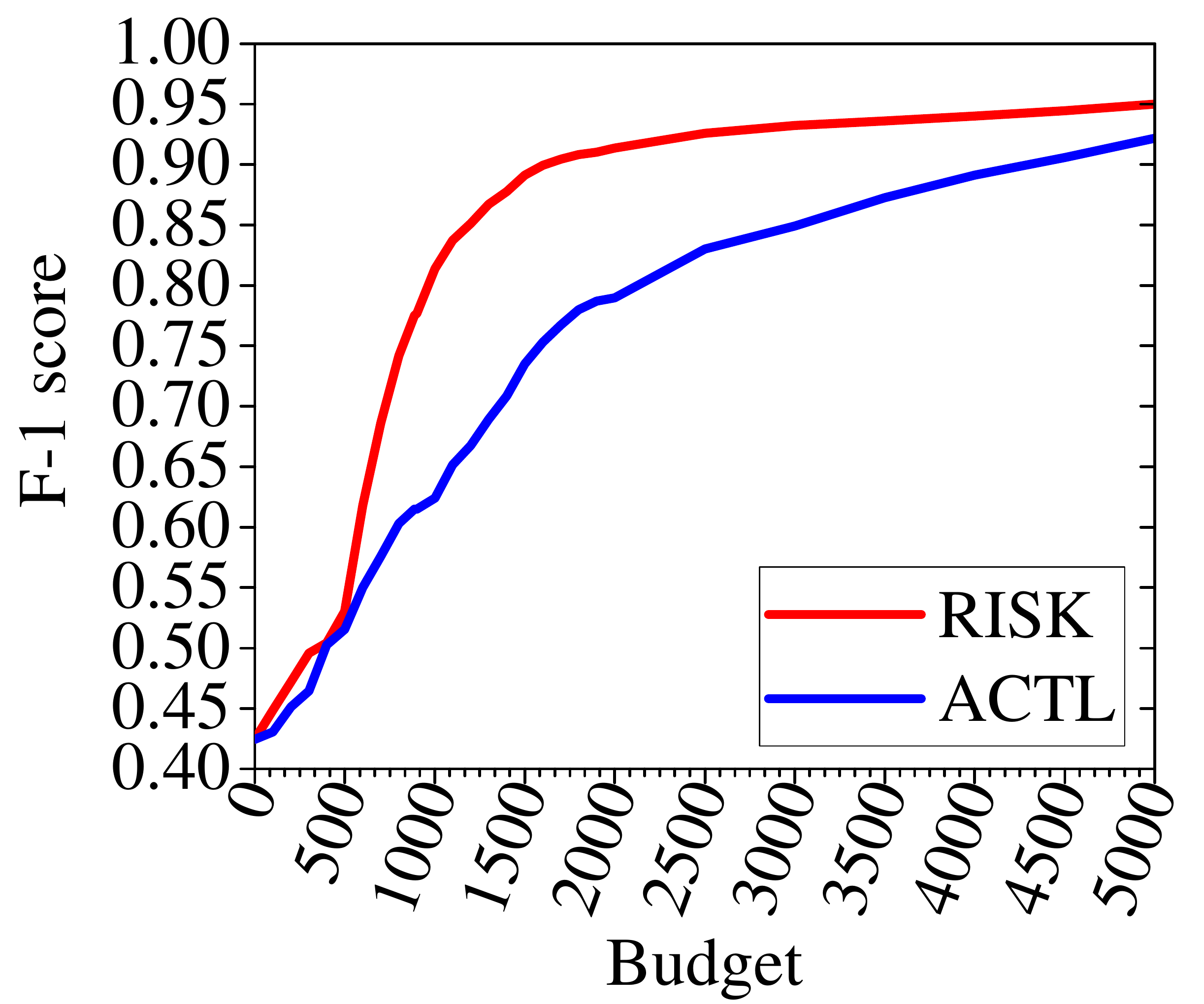}}
\vspace{-0.5cm}
\caption{Resolution quality comparison between RISK and ACTL.}
\label{fig:f1-score}
\vspace{-0.4cm}
\end{figure}
		
   The comparative results on pick-up accuracy are presented in Figure~\ref{fig:pairs-number}. It can be observed that provided with the same amount of budget, CVAR consistently picks up more mislabeled pairs than BASE and UNCT. Since both BASE and UNCT reason about the risk based on the match expectation estimated by the machine, it should not be surprising that they perform similarly. The improvement margins of CVAR over the alternatives first enlarge with the increase of budget, but then gradually narrow down as expected. Since the number of mislabeled pairs decreases with additional manual inspections, the performance difference between different approaches tend to decrease as well. These experimental results clearly validate the efficacy of the proposed risk model.
	
	The comparative results on resolution quality, measured by the F-1 metric, between RISK and ACTL, are also presented in Figure~\ref{fig:f1-score}. The achieved quality is measured on the results consisting of both manually labeled pairs and the pairs labeled by the classifier. It can be observed that after initial iterations, RISK achieves considerably better quality than ACTL. Even though ACTL uses the additional labeled data to update its classifier, the marginal benefit of additional training data points drops quickly with the increase of budget as expected. These experimental results show that the risk-based approach can be more effective than the active learning approach in improving resolution quality.

\vspace{-6pt}  \section{Conclusion} \label{sec:conclusion}
\vspace{-2pt}
  In this paper, we propose to investigate the problem of human and machine cooperation for ER from a risk perspective. We have presented a risk model and empirically validated its efficacy. It is worthy to point out that the proposed risk-based framework can be potentially generalized for other classification tasks. It is interesting to investigate its application in the scenarios besides ER in future work.   

\vspace{-6pt} \section*{Acknowledgment}
\vspace{-3.5pt}
This work was supported by the National Key R\&D Program of China (2016YFB1000703), NSF of China (61732014, 61332006, 61472321, 61502390 and 61672432) and Shaanxi NSBR Plan (2018JM6086).
\vspace{-8pt}
\bibliographystyle{ACM-Reference-Format}
\bibliography{references}

\end{document}